\documentclass[sigconf, nonacm]{acmart}

\usepackage{algorithm}
\usepackage{algpseudocode}
\usepackage{subcaption}

\newcommand\vldbyear{2024}
\newcommand\vldbworkshop{Foundations and Applications of Blockchain (FAB)}
\newcommand\vldbauthors{\authors}
\newcommand\vldbtitle{\shorttitle} 
\newcommand\vldbavailabilityurl{https://github.com/gdemos01/FAB-2024}
\newcommand\vldbpagestyle{plain} 

\begin{document}
\title{From On-chain to Macro: Assessing the Importance of Data Source Diversity in Cryptocurrency Market Forecasting}

%%
%% The "author" command and its associated commands are used to define the authors and their affiliations.
\author{Giorgos Demosthenous}
\affiliation{%
  \institution{University of Cyprus}
  \city{Nicosia}
  \state{Cyprus}
}
\email{demosthenous.giorgos@ucy.ac.cy}

\author{Chryssis Georgiou}
\affiliation{%
  \institution{University of Cyprus}
  \city{Nicosia}
  \state{Cyprus}
}
\email{georgiou.chryssis@ucy.ac.cy}

\author{Eliada Polydorou}
\affiliation{%
  \institution{University of Cyprus}
  \city{Nicosia}
  \state{Cyprus}
}
\email{polydorou.eliada@ucy.ac.cy}

\begin{abstract}
This study investigates the impact of data source diversity on the performance of cryptocurrency forecasting models by integrating various data categories, including technical indicators, on-chain metrics, sentiment and interest metrics, traditional market indices, and macroeconomic indicators. We introduce the Crypto100 index, representing the top 100 cryptocurrencies by market capitalization, and propose a novel feature reduction algorithm to identify the most impactful and resilient features from diverse data sources. Our comprehensive experiments demonstrate that data source diversity significantly enhances the predictive performance of forecasting models across different time horizons. Key findings include the paramount importance of on-chain metrics for both short-term and long-term predictions, the growing relevance of traditional market indices and macroeconomic indicators for longer-term forecasts, and substantial improvements in model accuracy when diverse data sources are utilized. These insights help demystify the short-term and long-term driving factors of the cryptocurrency market and lay the groundwork for developing more accurate and resilient forecasting models.
\end{abstract}

\maketitle

\pagestyle{\vldbpagestyle}
\begingroup\small\noindent\raggedright\textbf{VLDB Workshop Reference Format:}\\
\vldbauthors. \vldbtitle. VLDB \vldbyear\ Workshop: \vldbworkshop.\\ %\vldbvolume(\vldbissue): \vldbpages, \vldbyear.\\
%\href{https://doi.org/\vldbdoi}{doi:\vldbdoi}
\endgroup
\begingroup
\renewcommand\thefootnote{}\footnote{\noindent
This work is licensed under the Creative Commons BY-NC-ND 4.0 International License. Visit \url{https://creativecommons.org/licenses/by-nc-nd/4.0/} to view a copy of this license. For any use beyond those covered by this license, obtain permission by emailing \href{mailto:info@vldb.org}{info@vldb.org}. Copyright is held by the owner/author(s). Publication rights licensed to the VLDB Endowment. \\
\raggedright Proceedings of the VLDB Endowment. %, Vol. \vldbvolume, No. \vldbissue\ %
ISSN 2150-8097. \\
%\href{https://doi.org/\vldbdoi}{doi:\vldbdoi} \\
}\addtocounter{footnote}{-1}\endgroup

\ifdefempty{\vldbavailabilityurl}{}{
\vspace{.3cm}
\begingroup\small\noindent\raggedright\textbf{VLDB Workshop Artifact Availability:}\\
The source code, data, and/or other artifacts have been made available at \url{\vldbavailabilityurl}.
\endgroup
}

\section{Introduction}

Emerging in the catastrophic aftermath of the 2008 financial crisis \cite{schwartz2009origins}, which eroded trust in traditional financial institutions, Bitcoin not only challenged the established financial infrastructure but also pioneered the first pure digital asset class \cite{satoshi}, paving the way for what would eventually become a flourishing industry. The cryptocurrency market grew exponentially in recent years, with now more than 10000 assets circulating \cite{coingecko_2024}. This market has experienced rapid mainstream adoption, with an estimated 580 million people holding cryptocurrency by the end of 2023, marking a 34\% year-over-year increase \cite{CryptoCom2023}. The unique attributes and operational dynamics of cryptocurrencies, such as 24/7 trading availability, peer-to-peer transactions, minimal transaction fees, borderless nature, and enhanced privacy, make them particularly attractive to all levels of investors, from retail to institutional.

Despite the rising participation in cryptocurrency markets and the increasing volume of work around modeling and trading crypto assets \cite{fang2022cryptocurrency}, building accurate and resilient forecasting models for this market remains a real challenge. This can be attributed to the non-linear, highly volatile, uncertain, and noisy nature of the cryptocurrency market. The low volume of historical data surrounding this market also contributes to the problem, further highlighting the need to search for alternative information avenues. Most literature surrounding forecasting in cryptocurrency markets, focuses on incorporating technical indicators (e.g., OHLC candles) and sentiment analysis (e.g., social media sentiment) to build models capable of predicting market movements \cite{alonso2020convolution, ortu2022technical}. This approach is problematic because not only it underestimates the noise in this market and the necessity for additional, more valuable data sources, but also does not take advantage of data native to this market. 

On-chain data is a new family of metrics extracted from the underlying blockchain technology of cryptocurrencies, which offers a unique insight into investor behavior, network activity, and the supply dynamics of assets in this market. Realizing the importance of on-chain data, recent work has been incorporating it more and more in building their forecasting models and for better understanding how this market operates \cite{kim2021predicting, sebastiao2021forecasting, casella2023predicting}. In addition to technical, on-chain and sentiment metrics, there are other categories of data (e.g., macroeconomic and traditional market metrics), that when combined could further aid in deriving valuable information from the market, reducing noise, and consequently improving the performance of forecasting models. 

Paradoxically, most literature only combines a couple of data categories, usually chosen through intuition and without any prior experimentation (e.g.,~\cite{ortu2022technical, kim2021predicting}), which might severely undermine the potential of their forecasting models. In this work, we focus on analyzing a variety of different data categories and showcasing the impact that data source diversity can have on the performance of cryptocurrency forecasting models. We also investigate how the predictive power of each data category, as well as individual indicators, changes as prediction difficulty increases, laying the groundwork for better understanding this market and consequently building more accurate and resilient forecasting models, capable of both short-term and long-term predictions. Specifically, our {\bf contributions} are the following:\\
%\begin{itemize}
    \noindent\textbf{Crypto100 index:} We introduce the Crypto100 index, a new market index that tracks the top 100 cryptocurrencies by market capitalization. This index was designed to be directly comparable with any other asset in this market, and a good representation of the entire cryptocurrency market.\\     
    \noindent\textbf{Feature reduction algorithm:} We present a novel algorithm for creating feature vectors consisting of the most impactful and persistent features across a variety of complementary importance evaluation methods. This methodology ensures the identification of the most important and resilient features from a diverse set of data sources, enhancing model accuracy.\\
    \noindent\textbf{Impact analysis on data source diversity:} Our comprehensive experiments illustrate the significant influence of data source diversity on the predictive performance of forecasting models. We examine various time horizons and scenarios, including short-term (1 and 7 days) and long-term (90 and 180 days) prediction windows, highlighting the changing importance of different data categories and individual indicators across these periods.
%\end{itemize}

\section{Background}

\subsection{Forecasting Models}

Forecasting models are essential tools for predicting future market behavior based on historical data and current market inputs. These models, which include time series and regression analysis, econometric models, and machine learning algorithms, are crucial for informed decision-making, risk management, resource allocation, and performance evaluation in investment strategies \cite{fang2022cryptocurrency}. An additional benefit of building these models is their ability to extract useful information from noisy markets, leading to a better understanding of the factors driving market movements. In the context of cryptocurrency markets, forecasting models are most commonly used for predicting price movements and are integrated into trading strategies \cite{ren2022past}. In recent years, these models have also been employed to build more resilient cryptocurrency portfolios~\cite{ma2020portfolio}.

\subsection{Categories of Data Sources}

Different categories of data, ranging from macroeconomic indicators to on-chain metrics, offer distinct and diverse perspectives on the market, each providing unique insights into market behavior. In the context of this study, we have separated market information into the following categories:

%\begin{itemize}
\noindent\textbf{Macroeconomic Indicators}: These indicators encompass broader economic metrics and global financial data that can influence a variety of markets, including the cryptocurrency market. Macro indicators help contextualize cryptocurrency market movements within the larger economic climate. Examples of these indicators include central bank interest rates, inflation rates, policy uncertainty indices, and others.\\
\noindent\textbf{Technical Indicators}: Derived from the technical analysis of historical market information such as price and volume, these indicators aid in identifying patterns within the raw market data. Examples of technical indicators include moving averages, the relative strength index (RSI), and Bollinger Bands\\
\noindent\textbf{Sentiment and Interest Metrics}: This category aims to capture the mood, opinions, and interest of market participants from various sources like social media, news articles, and search trends. Shifts in investor sentiment can provide insights into the upcoming state of the markets. Common metrics include social media posts volume, google trends, social media post sentiments (e.g., negative, neutral, positive), and the fear and greed index.\\
\noindent\textbf{Traditional Market Indices}: These are major financial indices from traditional finance that can provide insights into a variety of markets as well as broader economic conditions. These indices include bonds (e.g., BSV \cite{vanguard_bsv_etf}), metals (e.g., gold), exchange rates (e.g., EURUSD), currency strength (e.g., U.S. Dollar index \cite{invesco_uup_etf}) and stocks (e.g., Nasdaq-100 index - QQQ).\\
\noindent\textbf{On-chain Metrics}: This new family of data points was introduced with the emergence of blockchain technology. These metrics are derived directly from the blockchain and offer unprecedented insights into market behavior by monitoring network activity and statistics. Examples of on-chain metrics include transaction volume, miner revenue, hash rate, active addresses, token distributions, supply information, network fees, and others.
%\end{itemize}

\section{Methodology}

\subsection{Datasets}

To assess the importance of data source diversity, historical data were collected and incorporated into various datasets used to evaluate the significance of individual metrics under different scenarios. For each data category, daily data were collected from sources such as Coinmetrics.io \cite{coinmetrics}, CoinGecko.com \cite{coingecko_2024}, the European Central Bank (ECB) \cite{ecb_dataset}, and others \cite{lunarcrush, policy_uncertainty}. Technical indicators were constructed using only BTC historical market information, as we found them to be highly correlated with and influential on the broader cryptocurrency market. Building on this observation, it was assumed and later validated (Subsection \ref{subsec: data_category_contribution}) that Bitcoin's on-chain metrics could also affect the rest of the market, and were therefore included in the dataset as market representatives. To investigate the importance of stablecoins, on-chain information from the second largest stablecoin by market capitalization, USDC, were also gathered. Although USDT has a much higher marketcap we chose not to include it due to it's bad reputation and questionable activity \cite{griffin2020bitcoin}, which might negatively influence this study leading to misleading conclusions. To collect on-chain information, we used Coinmetric's free API instead of running nodes on the blockchain networks to manually collect transaction data and calculate the respective metrics. Although this approach is easier and less costly, future research that requires more advanced on-chain metrics might need to consider directly running nodes on the networks instead. The data collection period spanned from January 2017 to June 2023.

\subsubsection{The Crypto100 index}

In contrast with most previous work which focuses solely on Bitcoin, Ethereum or a minor collection of popular cryptocurrencies, this study aims to capture a better representation of the entire crypto market. To achieve this, the Crypto100 index was created. This index tracks the top 100 cryptocurrencies by market capitalization, on a daily basis, and is calculated by the following equation:  

\begin{displaymath}
  \text{Crypto100} = \frac{\sum_{i=1}^{100} \text{Market Cap}_i}{\left(\log_{10}\left(\sum_{i=1}^{100} \text{Market Cap}_i\right)\right)^7}
\end{displaymath}
\vspace{0.5em} 

\begin{figure}[h]
  \centering
  \includegraphics[width=\linewidth]{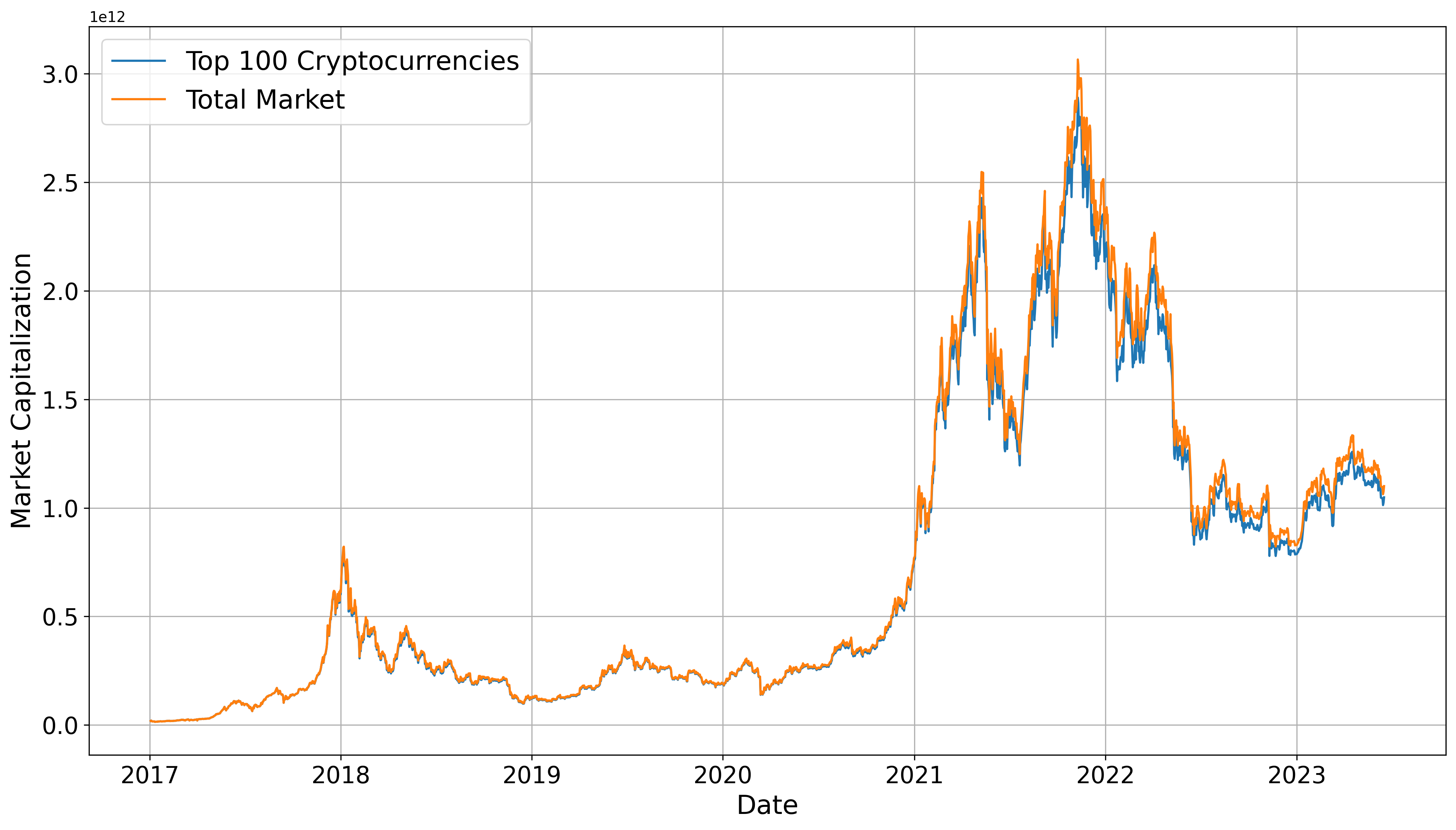}
  \caption{Top 100 Cryptocurrencies VS Total Marketcap}
  \Description{Top 100 Cryptocurrencies VS Total Marketcap}
  \label{fig:cr100_marketcap}
  % \vspace{-1.5em} 
\end{figure}

\begin{figure}[h]
    \centering
    % First image
    \begin{subfigure}[b]{\linewidth}
        \centering
        \includegraphics[width=\linewidth]{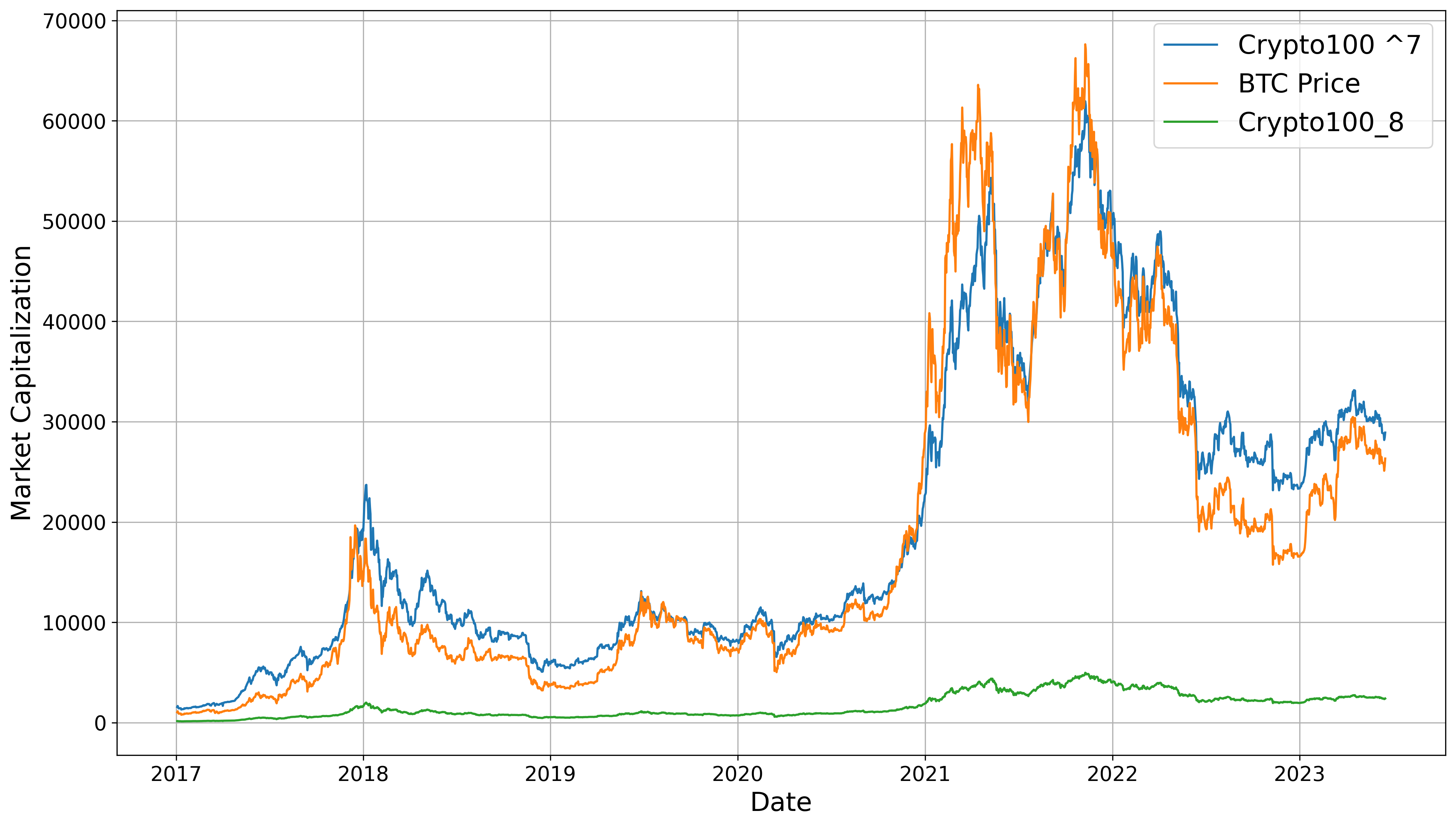}
        \caption{Scaling factor raised to the 7th and 8th power vs BTC price}
        \Description{A comparison between two versions of the Crypto100 index, using different scaling factors, against the Bitcoin price.}
        \label{fig:cr100_sub1}
    \end{subfigure}
    
    % Space between the two images
    \vspace{0.5cm}
    
    % Second image
    \begin{subfigure}[b]{\linewidth}
        \centering
        \includegraphics[width=\linewidth]{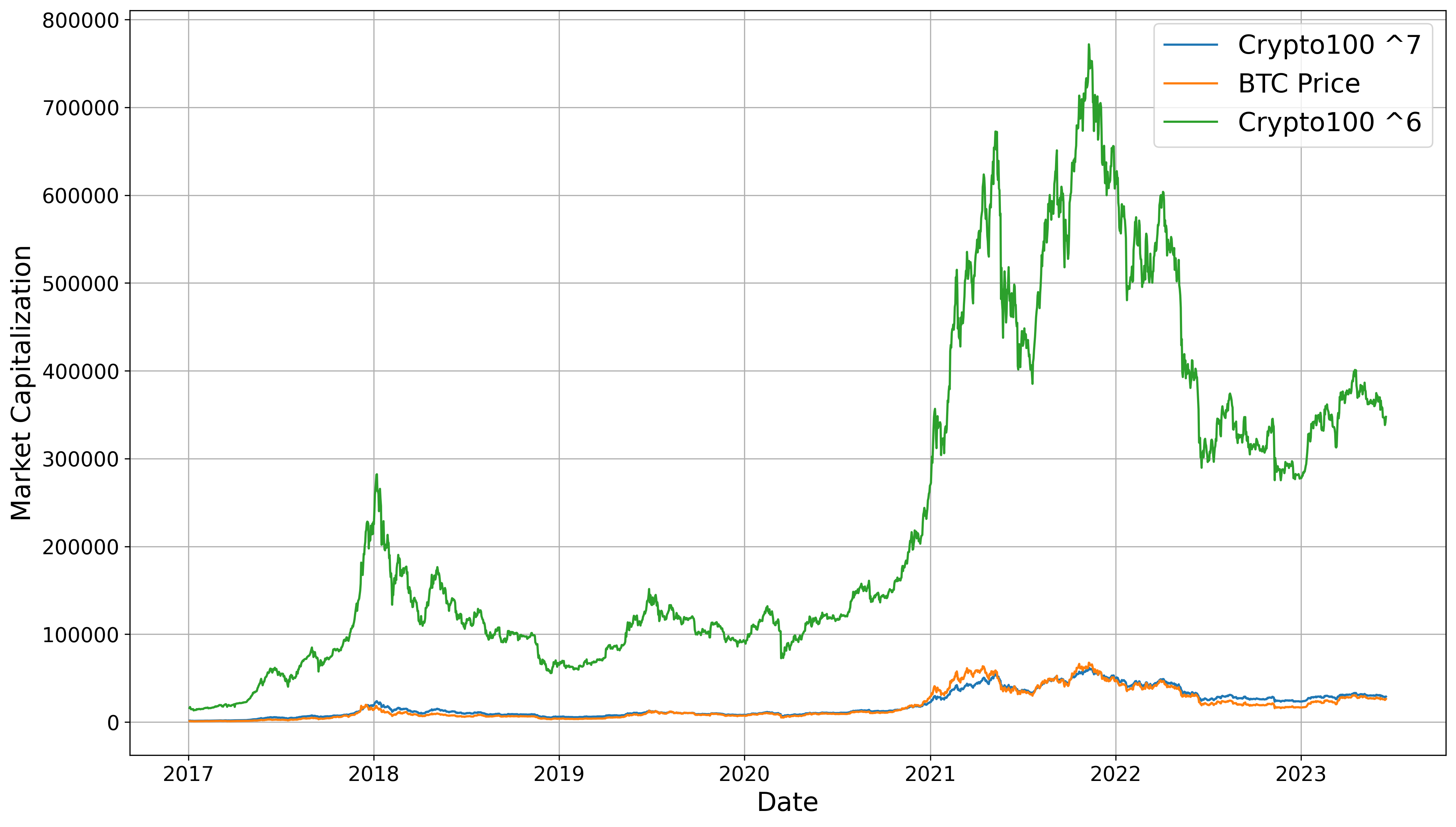}
        \caption{Scaling factor raised to the 6th and 7th power vs BTC price}
        \label{fig:cr100_sub2}
    \end{subfigure}
    
    \caption{Comparison of Crypto100 versions using different scaling factors with Bitcoin's price}
    \Description{Pending}
    \label{fig:crypto_comparison}
    % \vspace{-2em} 
\end{figure}

The top 100 cryptocurrencies constitute the majority of the market, as evidenced by their cumulative market capitalization (cf.~Figure~\ref{fig:cr100_marketcap}), making the Crypto100 index a good representation of the entire cryptocurrency market. Inspired by the calculation methodology of the S\&P500 index, which uses a divisor to normalize the final value and maintain index continuity over time~\cite{investopediasp500calculation}, Crypto100 employs a scaling factor derived from extensive experimentation. This approach ensures the Crypto100 index maintains continuity and resilience to sudden changes in the assets comprising the top 100 list, which is common in a maturing market like the cryptocurrency market. To make the Crypto100 index price directly comparable to other assets in the cryptocurrency market, we used Bitcoin's price to tune the scaling factor. Figure \ref{fig:cr100_sub2} shows that when the scaling factor is raised to powers lower than 6, it significantly limits its impact, making the Crypto100 index less comparable to other assets, such as Bitcoin. Thus, as illustrated in Figure \ref{fig:cr100_sub1}, the chosen power for tuning the scaling factor ensures the Crypto100 index is appropriately scaled and comparable to other assets in the market. For the index to maintain its resilience in this evolving market as time passes, the scaling factor should be reevaluated periodically and adjusted dynamically if necessary.

\subsubsection{Scenarios}

A significant challenge in a maturing market, such as the cryptocurrency market, is that experiments conducted over different chronological periods can yield varying results \cite{mokni2024efficiency}. Bearing this in mind, alongside investigating the entire 2017-2023 period, a subset of the original dataset starting from January 2019 was also created. The rationale for selecting January 2019 as the starting date for the subset is two-fold: Firstly, this month marked the bottom of the previous bull market, but more importantly, many critical data points, such as the USDC stablecoin and the fear-and-greed index, were only introduced after the end of 2018.

To examine how data source diversity impacts forecasting models across short-term and long-term horizons, fine-tuned machine learning (ML) models were trained to predict the Crypto100 index price 1, 7, 30, 90, and 180 days in advance, which are \textit{prediction windows} commonly found in literature \cite{ren2022past, fang2022cryptocurrency}. The importance of individual indicators for each prediction window was also extracted and analyzed. Initially, the datasets consisted of 429 metrics across five categories, with on-chain data separated into two subcategories: \textit{On-chain\_BTC} and \textit{On-chain\_USDC}. Following an initial data cleaning and preprocessing phase that included the standard methods used in ML \cite{chicco2022eleven} such as filling empty data with interpolation, removing duplicate values, and discarding features that had flat or missing values for very long periods, two new subsets were created from the original dataset: \textit{set 2017}, covering the period between January 2017 and June 2023, and \textit{set 2019}, covering the period between January 2019 and June 2023. Metrics that began recording their values after the initial date of each period (e.g., USDC starting in 2018), were discarded from the corresponding set. As a result, the 2017 and 2019 sets comprised 192 and 283 metrics, respectively.%\vspace{-1em}

\subsection{Feature Selection}
To narrow down the focus of the investigation to the most important indicators for each scenario, a robust feature selection methodology was devised, combining {\em four} different methods: Pearson correlation, Mean Decrease Impurity (MDI), Permutation Feature Importance (PFI), and SHapley Additive exPlanations (SHAP) \cite{lundberg2017unified}. Although the correlation between features and the target variable provides a good initial indication of the importance of individual features, it cannot be used as a standalone method since it fails to capture non-linear relationships. Therefore, feature selection algorithms based on machine learning models were also incorporated into our methodology. Specifically, Random Forests (RF) and XGBoost (XGB) were fine-tuned using 5-fold cross-validation grid search \cite{scikit-learn_grid_search} with minimum mean squared error (MSE) as the objective for each of the 10 different scenarios (2 sets x 5 prediction windows). The search space included parameters relevant to tree structures like number of estimators, maximum depth, sample splits, etc. These models were then used to calculate feature importance using their built-in MDI-based algorithm. Additionally, PFI was also extracted for both models using MSE as the optimization measure. In contrast to MDI, PFI directly measures the effect on each model's predictive performance, mitigating issues caused by bias during training.

The Feature Reduction Algorithm (Algorithm \ref{algo:FRA}) is a novel methodology for creating feature vectors consisting of the most impactful and persistent features across a variety of different importance evaluation methods. The objective of the algorithm is to iteratively remove features that consistently rank at the bottom of each inner evaluation method until the dimensionality of the vector is reduced below the desired number. Due to the low volume of records in our dataset, we needed to substantially reduce the feature vector's length to help models distinguish valuable information from noise. To retain important indicators, we set the target length at 100, ensuring we could still examine the feature vector comprehensively, as significant insights might be present throughout, including the tail end. The inner methods used are MDI and PFI with two different models (RF and XGB), as well as Pearson correlation. The correlation value, initially set to 0.5, is used as a threshold for removal of surviving bottom features and increases with each iteration by 0.025 to make the selection process stricter until the algorithm's objective is met. The strict approach using complementary importance evaluation methods of the Feature Reduction Algorithm (FRA) ensures that each method's strengths are leveraged while mitigating their individual weaknesses, thereby limiting the possibility of noisy features appearing in the final selection.

\begin{algorithm}
\caption{Feature Reduction Algorithm (FRA) for selecting the most important indicators across all categories}
\label{algo:FRA}
\begin{algorithmic}[1]
\State periods $\gets$ [2017, 2019]
\State prediction\_window $\gets$ [1, 7, 30, 90, 180]
\For{each period}
    \For{each prediction window}
        \State $corr\_threshold \gets 0.5$
        \While{$number\_of\_features > 100$}
            \State $RF\_FI \gets$ Extract importance using RF (MDI)
            \State $XGB\_FI \gets$ Extract importance using XGB (MDI)
            \State $RF\_PFI \gets$ Extract importance using RF (MSE)
            \State $XGB\_PFI \gets$ Extract importance using XGB (MSE)
            \If {$feature\_rank$ = $bottom\_50\%$ \textbf{AND} $feature\_cor$ < $corr\_threshold$}
                \State Remove corresponding features
            \EndIf
            \State $corr\_threshold \gets corr\_threshold + 0.025$
        \EndWhile
        \State \Return reduced feature vector
    \EndFor
\EndFor
\end{algorithmic}
\end{algorithm}

To validate the efficacy of FRA and to mitigate possible drawbacks of the algorithm, SHAP was used to compute the contribution of individual features from the original sets, and combine its output with FRA to export the final feature vector. Although SHAP was originally used as a method to help interpret the output of machine learning models, in recent years it has been gaining popularity mainly as a feature importance evaluation approach. It works by computing the Shapley value for each feature, which is derived from cooperative game theory and represents the average contribution of a specific feature across all possible combinations of features. The overlap between the top 100 features extracted by SHAP and the reduced feature vector of FRA (< 100) is on average 78 features, which further validates that the surviving features are indeed the most important ones. To create the final feature vector for each scenario (Table \ref{tab:feature_vectors}), we take the union of the top 75 features ranked by importance from FRA and SHAP.

\begin{table}
\small
    \caption{Summary of final feature vectors for all different scenarios (year\_prediction window)}
    \label{tab:feature_vectors}
    \begin{tabular}{rc}
        \toprule
        Scenario & Number of Features \\
        \midrule
        2017\_1 & 79 \\
        2017\_7 & 79 \\
        2017\_30 & 81 \\
        2017\_90 & 86 \\
        2017\_180 & 88 \\
        2019\_1 & 100 \\
        2019\_7 & 97 \\
        2019\_30 & 100 \\
        2019\_90 & 91 \\
        2019\_180 & 90 \\
        \bottomrule
    \end{tabular}%\vspace{-2em}
\end{table}

% shap 2017: 91, 87, 88, 83, 84 - 86.6 ~ 87
% shap 2019: 71, 67, 63, 71, 73 - 69

\subsection{Impact Demonstration}

Several approaches were followed to demonstrate the impact of data source diversity. Initially, the contribution of individual data categories to the final feature vector is analyzed across all prediction windows to discover how it changes over time (Section \ref{subsec: data_category_contribution}). We then group scenarios together into short-term and long-term time horizons and proceed to analyze both the top and unique indicators found in each use case (Section \ref{subsec: shor_long_term_factors}). To investigate the explicit effect of data source diversity on forecasting models, we fine-tune and train ML models for each scenario and explore how their performance changes when using the diverse feature vector compared to only using each individual category (Section \ref{subsec: model_perfomance_improvement}).

\section{Results}

\subsection{The Contribution of Individual Data Sources}
\label{subsec: data_category_contribution}

\begin{figure} [h]
  \centering
  \includegraphics[width=\linewidth]{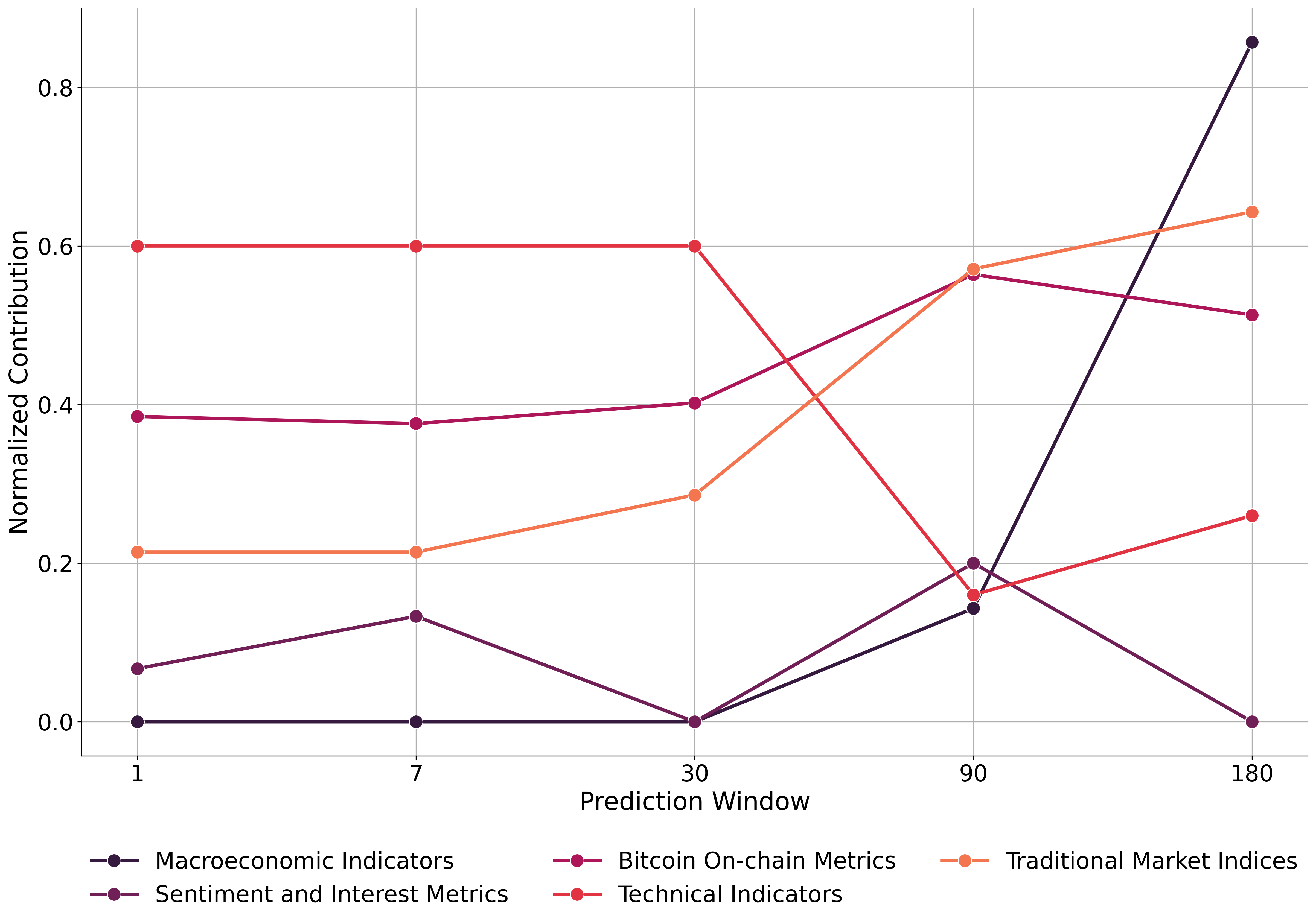}
  \caption{Contribution of individual data sources to the final feature vector in set 2017 across all prediction windows}
  \Description{Contribution of each data category to the final feature vector for the set 2017, analyzed across prediction windows of 1, 7, 30, 90, and 180 days.}
  \label{fig:contribution_2017}
  %\vspace{-2em} 
\end{figure}

\begin{figure} [h]
  \centering
  \includegraphics[width=\linewidth]{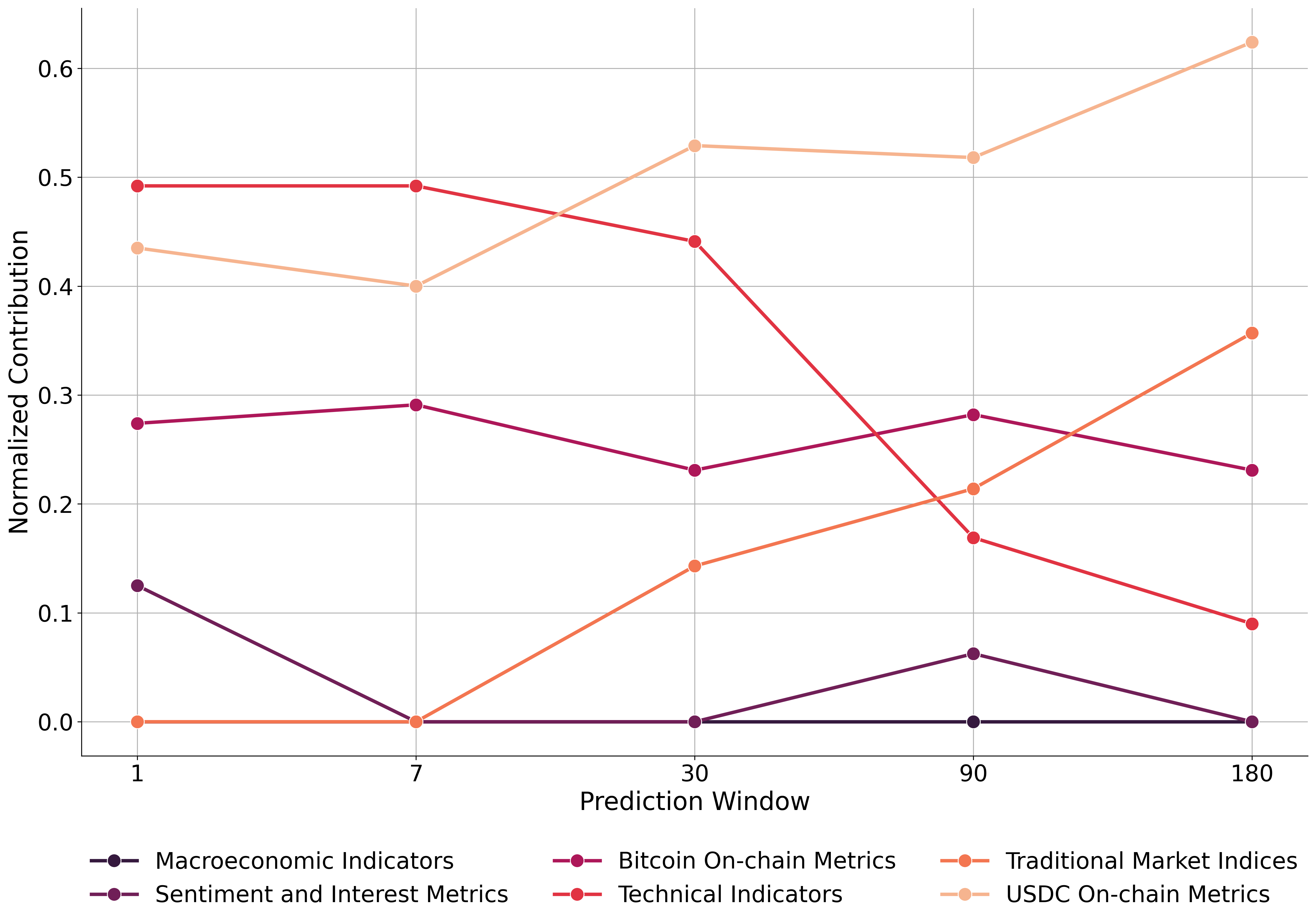}
  \caption{Contribution of individual data sources to the final feature vector in set 2019 across all prediction windows}
  \Description{Contribution of each data category to the final feature vector for the set 2019, analyzed across prediction windows of 1, 7, 30, 90, and 180 days.}
  \label{fig:contribution_2019}
  %\vspace{-1em} 
\end{figure}

The importance of an individual data category in forecasting models can vary significantly depending on the prediction window, with the impact of different indicators changing as the time horizon moves from short-term to long-term predictions. To better understand this changing dynamic between indicator categories, the contribution of each data source to the final feature vector, extracted using the FRA algorithm, across all prediction windows was recorded for both set 2017 (Figure \ref{fig:contribution_2017}) and set 2019 (Figure \ref{fig:contribution_2019}). To make the categories comparable, we calculate their contribution by dividing the final number of features from the category included in the final vector with the corresponding total number of candidate features in the same category before the feature selection phase took place. In other words, the {\em contribution factor} is the ratio of features from each category included in the final feature vector.

Looking at Figures \ref{fig:contribution_2017} and \ref{fig:contribution_2019} it immediately becomes apparent that \textit{On-chain Metrics} maintain a very high contribution to the final feature vector across all prediction windows. This behavior demonstrates that this new data family not only seems to be very important for cryptocurrency forecasting models, but also appears to be suitable for short-, medium- and long-term predictions. 

\textit{Technical Indicators} on the other hand appear to be very valuable for short-term predictions, but their standalone predictive power quickly diminishes over longer-periods. Their ability to capture evolving market patterns and trends makes them essential for short-term windows, but their decreasing contribution over longer windows suggests that technical indicators need to be supplemented with other data categories to build models capable of long-term forecasting. 

Although the contribution of \textit{Traditional Market Indices} is initially at low levels, it rapidly increases for longer prediction windows, eventually becoming the second-highest contributor of indicators to the final feature vector in both 2017 and 2019 sets. The growing relevance of traditional market indices over longer periods highlights the increasing integration and correlation between traditional financial markets and cryptocurrencies. As the cryptocurrency market matures, it becomes more sensitive to broader market conditions and trends, making traditional indices more relevant for long-term predictions. 

As depicted in Figure \ref{fig:contribution_2017}, \textit{Macroeconomic Indicators} appear to have no effect in short-term predictions, but their contribution rises in long-term prediction windows. This can be attributed to the delayed effect of economic policies and other macroeconomic factors that gradually permeate all financial markets, including cryptocurrencies. Paradoxically, macroeconomic indicators are completely absent from the 2019 set (Figure \ref{fig:contribution_2019}), which might be due to any of the following reasons: $(i)$ the decreased size of the 2019 datasets doesn't include significant changes in macroeconomic indicators, which usually become evident over longer time horizons, $(ii)$ the cryptocurrency market in certain time periods (e.g., periods of rapid increase of mainstream adoption), might become more self-contained and independent of broader economic conditions compared to other markets, and $(iii)$ the dominance of other factors such as the inclusion of USDC metrics might simply overshadow macroeconomic influences. The discrepancies observed between the 2017 and 2019 sets further validate the assumption that studying different periods in the cryptocurrency market can sometimes yield varying results. To confirm that these differences are due to changing market behavior and not noise, future research could focus on enhancing FRA by incorporating more dynamic elements, thereby increasing its robustness.

The introduction of stablecoins such as USDC to the cryptocurrency markets aided in increasing market stability and liquidity \cite{catalini2022some}. Naturally, \textit{On-chain Metrics (USDC)} have a very important role to play in forecasting market movements across all time horizons (Figure \ref{fig:contribution_2019}). Their contribution substantially surpasses even that of Bitcoin's on-chain data, becoming by far the most important contributor of indicators to the final feature vector for 30, 90 and 180 day prediction windows. The reason for this could be that USDC data encapsulate information about macro changes of the crypto market, as well as market sentiment and investor behavior. Inflow/outflow information of the USDC stablecoin can be good indicators of changes in the buying/selling behaviors of investors, profit taking periods, moving funds in and out of the cryptocurrency market, etc. Because of the increasing importance of USDC-based on-chain data and the evolving impact of stablecoins on the cryptocurrency market, further research into the dynamics of individual indicators within this category is needed.

\textit{Sentiment and Interest Metrics} have a high contribution for short-term predictions, with their effect diminishing on average over longer periods. This could be attributed to immediate market reactions driven by social media, news, and investor interest, which can cause short-term price changes. Although some indicators in this category, like monthly search volume measured by Google trends, provide intelligence about longer time horizons (hence the increased contribution for 90-day window), the majority supply information that can drive short-term price volatility but cannot sustain long-term trends.

\subsection{Short-term and Long-term Driving Factors}
\label{subsec: shor_long_term_factors}

To delve deeper into which individual metrics affect the predictive capabilities of forecasting models, two groups were created based on the final feature vectors. Specifically, features from sets with different prediction windows were merged together to form the \textit{Short-term} (windows 1 and 7) and \textit{Long-term} (windows 90 and 180) groups. Fine-tuned random forests were used to extract an importance value for each feature in the final feature vectors. When creating the groups, the importance of common features within a group was calculated by averaging their importance value before merging. To better understand which individual indicators affect each group and why, their brief definitions are provided in Table \ref{tab:fet_definitions}.

\begin{table} %[h!]
\small
    \caption{Definitions of the features used in our analysis}
    \label{tab:fet_definitions}
    \begin{tabular}{p{0.33\linewidth} p{0.6\linewidth}}
        \toprule
        Feature & Definition \\
        \midrule
        % Address Balance Features
        (usdc)\_AdrBal1in\#Cnt & Sum count of unique addresses holding at least one \textit{\#th} ownership of the current supply of USDC/BTC \\
        (usdc)\_AdrBalUSD\#Cnt & Sum count of unique addresses holding at least \textit{\$\#} worth of USDC/BTC) \\
        (usdc)\_AdrBalNtv\#Cnt & Sum count of unique addresses holding at least \textit{\#} USDC/BTC \\
        
        % Supply Features
        (usdc)\_SplyActPct\#yr & Percentage of the current USDC/BTC supply that has been active in the trailing \textit{\#} year \\
        (usdc)\_SplyActEver & Sum of unique USDC/BTC held by accounts that transacted at least once \\
        (usdc)\_SplyAdrBalNtv\# & Sum of all USDC/BTC being held in addresses whose balance is \textit{\#} native units or greater \\
        (usdc)\_SplyAdrBalUSD\# & Sum of all USDC/BTC being held in addresses whose balance is \textit{\$\#} or greater \\
        (usdc)\_SplyCur & The current supply of USDC/BTC \\
        s2f\_ratio & The stock-to-flow ratio of BTC \\
        
        % Capitalization Features
        (usdc)\_CapAct1yrUSD & Sum USD value of all active USDC/BTC in the last year \\
        CapRealUSD & The sum USD value based on the USD closing price on the day that a native unit last moved \\
        (usdc)\_CapMrktFFUSD & Sum USD value of the current free float supply(USDC/BTC) \\
        market\_cap & The market capitalization of BTC \\
        
        % Moving Average Features
        EMA\#\_variable & The exponential moving average of a \textit{variable} based on the \textit{\#} past days \\
        SMA\#\_variable & The simple moving average of a \textit{variable} based on the \textit{\#} past days \\
        
        % Wallet Size Features
        fish\_pct & The ratio of wallets holding 10-100 BTC \\
        shrimps\_pct & The ratio of wallets holding less than 10 BTC \\
         total\_balance & Sum of all BTC being held in addresses owned by whales, sharks, fish or shrimp \\
        
        % Google Trends Features
        gt\_X\_monthly & The monthly search volume of \textit{X} in Google trends \\
        
        % Price and Revenue Features
        X\_Close & The close price of index/indicator \textit{X} \\
        RevAllTimeUSD & Sum USD value of all Bitcoin miner revenue (fees plus newly issued BTC) since genesis \\
        RevHashRateUSD & The USD value of the mean daily miner reward per estimated hash unit per second performed during the day \\

        % Supply Equality Features
        SER & Supply Equality Ratio. The ratio of supply held by addresses with less than one ten-millionth of the current supply of USDC/BTC to the supply held by the top one percent of addresses. \\
        
        % Miner and Balance Features
        SplyMiner0HopAllUSD & The sum of the balances of all mining entities in USD \\

        % Velocity Features
        VelCur1yr & The ratio of the value transferred in the trailing 1 year divided by the current supply \\
        \bottomrule
    \end{tabular}%\vspace{-2em}
\end{table}

The \textit{top 5 features} of each group and for each set, ranked by importance, can be seen in Table \ref{tab:top5_features}. Following market trends (e.g., through moving averages) seems to be very important for short-term predictions (EMA100/EMA200). A possible reason for this, is that they provide a clear insight about the overall market direction which, when combined with other indicators, can improve prediction performance. The all-time revenue generated by miners (RevAllTimeUSD) is important across all scenarios, present in almost all feature vectors, and even ranks amongst the most important features for short-term and long-term predictions in sets 2017 and 2019 respectively. High miner revenue can signal strong network usage, which not only increases value and confidence in the cryptocurrency market but can also potentially influence price movements. Realized market capitalization (CapRealUSD) provides a better insight into the fair value of a crypto asset based on the actual transaction history rather than the current price, hence the observed increased importance. This indicator shows how much of Bitcoin's supply has been moved recently offering clues about how the market values the asset and where the price is directed. Although in the 2017 set retail investors are very important (AdrBalUSD100Cnt) in the short-term, medium and larger holders (fish\_pct, SplyAdrBalUSD10K, usdc\_AdrBalNtv10KCnt) also seem to have growing influence on short-term market movements based on the 2019 set, which could indicate that participation by wealthier and more accredited investors has increased in recent years. Looking at the top features in the Long-term group, it becomes clear that on-chain metrics based on balance and supply dynamics of the market have a lot of predictive power. Wealth distribution among smaller holders (SplyAdrBalUSD100, SplyAdrBalNtv0.01,SplyAdrBalUSD1, etc.) is suprisingly what drives longer-term market movements, despite the gradual increase of larger holders. Metrics about existing and active supply (SplyCur, SplyActEver) are also important long-term indicators because they encapsulate information about inflation and scarcity which do affect an asset's price in the longer timeframe.%\vspace{-.3em}

\begin{table}[h!]
\small
    \caption{Top 5 most important features for short-term and long-term Predictions in 2017 and 2019 sets}
    \label{tab:top5_features}
    \centering
    \begin{tabular}{lll}
        \toprule
        \textbf{Set} & \textbf{Short-term} & \textbf{Long-term} \\
        \midrule
        2017 & \begin{tabular}[c]{@{}l@{}} 
            EMA100\_market-cap \\ 
            EMA200\_close-price \\ 
            RevAllTimeUSD \\ 
            CapRealUSD \\ 
            AdrBalUSD100Cnt 
        \end{tabular} 
        & \begin{tabular}[c]{@{}l@{}} 
            SplyAdrBalUSD100 \\ 
            SplyAdrBalNtv0.01 \\ 
            SplyCur \\ 
            SplyAdrBalUSD1K \\ 
            SplyActEver 
        \end{tabular} \\
        \midrule
        2019 & \begin{tabular}[c]{@{}l@{}} 
            usdc\_AdrBalNtv1Cnt \\ 
            fish\_pct \\ 
            SplyAdrBalUSD10K \\ 
            usdc\_AdrBalNtv10KCnt \\ 
            total\_balance 
        \end{tabular} 
        & \begin{tabular}[c]{@{}l@{}} 
            AdrBalNtv0.1Cnt \\ 
            SplyAdrBalUSD10 \\ 
            RevAllTimeUSD \\ 
            SplyAdrBalNtv0.001 \\ 
            SplyAdrBalUSD1 
        \end{tabular} \\
        \bottomrule
    \end{tabular}%\vspace{-1em}
\end{table}

To investigate which features are uniquely important in each group, the 20 features with the highest importance that appeared in one group but not the other were selected and can be seen in Table \ref{tab:unique_top20}. Indicators around recent market movements dominate the unique driving factors of short-term forecasting markets. More recent SMAs/EMAs around price and market capitalization, with a past window of 5 to 30 trailing days, are the most populant in the Short-term group. Broader economic conditions evidenced by major indices appear to have a long-term effect in the predictability of the cryptocurrency market. These indexes include top non-financial companies (QQQ\_Close), the strength of the dollar (UUP\_close) and the euro (EURUSD\_Close), as well as the bond markets (BSV\_CLOSE, MBB\_close). The unique-feature analysis also further validates that supply dynamics and market activity metrics (SplyActPct1yr, SER, VelCur1yr, s2f\_ratio) can substantially affect long-term predictions. It is apparent that stablecoin data, in this case On-chain USDC metrics, have a major role to play in both short- and long-term predictions as evidenced by the sheer number of respective features present in the top 20 unique feature groups. In the short-term, USDC metrics focus on address balances (e.g., usdc\_AdrBalNtv10KCnt), which can indicate the recent activity and behavior of market participants, such as investors entering/leaving the market. In the long-term, USDC metrics highlight wealth distribution (e.g., usdc\_SplyAdrBalNtv100), supply activity (e.g., usdc\_SplyAct2yr), and supply dynamics (e.g., usdc\_SplyCur), which are indicators of long-term market stability and direction.

\begin{table}[h!]
\small
    \caption{Top 20 unique features for short-term and long-term predictions in 2017 and 2019 sets}
    \label{tab:unique_top20}
    \centering
    \begin{tabular}{lll}
        \toprule
        \textbf{Set} & \textbf{Short-term} & \textbf{Long-term} \\
        \midrule
        2017 & \begin{tabular}[c]{@{}l@{}} 
            AdrBalUSD100Cnt \\ 
            EMA14\_close-price \\ 
            EMA10\_close-price \\ 
            EMA10\_market-cap \\ 
            AdrBal1in1MCnt \\ 
            EMA20\_close-price \\ 
            EMA20\_market-cap \\ 
            EMA14\_market-cap \\ 
            market-cap \\ 
            EMA5\_market-cap \\ 
            SMA\_20\_close-price \\ 
            SMA\_10\_market-cap \\ 
            EMA5\_market-cap \\ 
            SMA\_20\_market-cap \\ 
            SMA\_10\_close-price \\ 
            EMA30\_close-price \\ 
            CapAct1yrUSD \\ 
            SMA\_5\_close-price \\ 
            SplyMiner0HopAllUSD \\ 
            EMA30\_market-cap 
        \end{tabular} 
        & \begin{tabular}[c]{@{}l@{}} 
            QQQ\_Close \\ 
            AdrBalNtv0.1Cnt \\ 
            AdrBal1in1BCnt \\ 
            AdrBalNtv0.01Cnt \\ 
            SplyActPct1yr \\ 
            AdrBal1in10KCnt \\ 
            UUP\_Close \\ 
            EMA200\_volume \\ 
            gt\_Ethereum\_monthly \\ 
            AdrBal1in10BCnt \\ 
            EURUSD\_Close \\ 
            SER \\ 
            EMA100\_volume \\ 
            RevHashRateUSD \\ 
            BSV\_Close \\ 
            VelCur1yr \\ 
            s2f\_ratio \\ 
            shrimps\_pct \\ 
            MBB\_Close \\ 
            SplyAdrBal1in1M 
        \end{tabular} \\
        \midrule
        2019 & \begin{tabular}[c]{@{}l@{}} 
            fish\_pct \\ 
            SplyAdrBalUSD10K \\ 
            usdc\_AdrBalNtv10KCnt \\ 
            usdc\_AdrBalUSD1MCnt \\ 
            usdc\_AdrBalUSD100KCnt \\ 
            AdrBalUSD100Cnt \\ 
            SplyAdrBalUSD10M \\ 
            usdc\_AdrBalUSD10KCnt \\ 
            usdc\_SplyAct7d \\ 
            AdrBalUSD1Cnt \\ 
            SplyAdrBal1in100K \\ 
            gt\_Cryptocurrency\_m \\ 
            SplyAdrTop1Pct \\ 
            EMA14\_close-price \\ 
            SplyAdrBalUSD1M \\ 
            SplyAdrBalNtv100 \\ 
            EMA10\_market-cap \\ 
            usdc\_SplyAdrBalUSD10M \\ 
            EMA20\_market-cap \\ 
            EMA14\_market-cap 
        \end{tabular} 
        & \begin{tabular}[c]{@{}l@{}} 
            AdrBalNtv0.1Cnt \\ 
            usdc\_SplyAdrBalNtv100 \\ 
            SplyActPct1yr \\ 
            usdc\_SplyAdrBalNtv0.001 \\ 
            usdc\_SplyCur \\ 
            UUP\_Close \\ 
            EURUSD\_Close \\ 
            SER \\ 
            usdc\_SplyAdrBal1in100M \\ 
            usdc\_SplyAct2yr \\ 
            SplyAct1yr \\ 
            gt\_Ethereum\_monthly \\ 
            usdc\_CapMrktFFUSD \\ 
            usdc\_SplyAdrBalUSD10 \\ 
            EMA200\_volume \\ 
            ROI1yr \\ 
            usdc\_SplyAdrBalNtv1 \\ 
            AdrBal1in1BCnt \\ 
            usdc\_SplyAct3yr 
        \end{tabular} \\
        \bottomrule
    \end{tabular}%\vspace{-2em}
\end{table}

\subsection{Model Performance Improvement}
\label{subsec: model_perfomance_improvement}

The results discussed up to this point have proven that combining metrics from different categories not only provides additional insight into important driving factors of short- and long-term market movements, but is also important in improving the predictive performance of forecasting models. To quantify the impact of data source diversity on the performance of forecasting models, ML models were fine-tuned and trained for each scenario. We then explored how their performance changes when using the diverse feature vector compared to only using data from a single category. {\em Performance improvement} is defined as the percentage decrease of the mean squared error after evaluating the model on the diverse feature vector. Tables \ref{tab:pw_improvement} and \ref{tab:dsc_improvement} include the average performance improvements of a RF model fine-tuned, using 5-fold cross-validation grid search, and trained for each respective scenario.

In Table \ref{tab:pw_improvement}, the average model improvement for each prediction window under both 2017 and 2019 sets is presented. With the notable exception of the shortest prediction window (w=1), the impact of data source diversity on the forecasting model seems to be steadily increasing over longer time periods. This is because predicting the future price of Crypto100, as well as any other price metric, naturally becomes more challenging as the prediction window increases, due to larger amounts of uncertainty regarding the future of the market, and therefore requires the additional insight that arises from the relationships between diverse data categories. One explanation about the shortest window exception could be that certain categories like \textit{Sentiment and Interest Metrics} as well as \textit{Technical Indicators} provide information which can mostly aid in predicting immediate market reactions, but the importance of which quickly fades in longer prediction timeframes as discussed in Section \ref{subsec: data_category_contribution}. Table \ref{tab:dsc_improvement} includes the average model improvements for each data category under both 2017 and 2019 sets. Categories that had the least representation in the final feature vector, as well as the initial dataset, such as \textit{Sentiment and Interest Metrics} and \textit{Macroeconomic Indicators}, benefited the most from data source diversity with the improvement level reaching all the way up to 1118.16\%. \textit{Bitcoin On-chain Metrics} on the other hand showcased the least improvement, which was 12.09\% and 17.51\% for 2017 and 2019 sets respectively. One reason for this could be that on-chain metrics include data more relevant to the cryptocurrency market, which capture a variety of both technical and fundamental information (investor activity, supply and demand dynamics, etc.), therefore making some of the other data categories less necessary.

Overall, the performance of the RF forecasting model improved on average by 455.67\% and 426.67\% in 2017 and 2019 sets, respectively. To validate whether this substantial improvement of performance is also observed in other models, an XGBoost estimator was also fine-tuned, trained and evaluated for every scenario. In the case of XGB, the average overall improvement for the 2017 set was 399.67\% and for the 2019 set was 468\%, which constitutes more evidence that data source diversity can have a major impact on the predictive power of cryptocurrency market forecasting models.%\vspace{.5em}

\begin{table}[t]
\small
    \caption{Average MSE percentage decrease of the RF model by prediction window for both 2017 and 2019 sets}
    \label{tab:pw_improvement}
    \begin{tabular}{rcl}
        \toprule
        & \multicolumn{2}{c}{Improvement (\%)} \\
        \cmidrule{2-3}
        Prediction Window & 2017 & 2019 \\
        % Pred. Window&2017 Improvement(\%)&2019 Improvement(\%)\\
        \midrule
        1 & 855.87\% & 794.71\%\\
        7 & 189.08\% & 191.51\%\\
        30 & 218.96\% & 274.74\%\\
        90 & 378.23\% & 359.81\%\\
        180 & 636.24\% & 512.59\%\\
        \bottomrule
    \end{tabular}%\vspace{-2em}
\end{table}

\begin{table}[t]
\small
    \caption{Average MSE percentage decrease of the RF model by data category for both 2017 and 2019 sets}
    \label{tab:dsc_improvement}
    \begin{tabular}{rcl}
        \toprule
        & \multicolumn{2}{c}{Improvement (\%)} \\
        \cmidrule{2-3}
        Prediction Window & 2017 & 2019 \\
        \midrule
        Macroeconomic Indicators &	825.72\%	&913.55\% \\
        Sentiment and Interest Metrics	&1118.16\%	&895.61\% \\
        On-chain Metrics (BTC)	&12.09\%	&17.51\% \\
        Traditional Market Indices	&244.38\%	&273.13\% \\
        Technical Indicators	&78.02\%	&81.72\% \\
        On-chain Metrics (USDC)	&-	&378.52\% \\
        \bottomrule
    \end{tabular}%\vspace{-2em}
\end{table}

\section{Challenges and future work}

\textbf{Balanced category representation}. A drawback of this study is that some data categories like \textit{Sentiment and Interest Metrics} and \textit{Macroeconomic Indicators} are underrepresented in the original dataset. This might have led to underestimating their importance in certain scenarios. In future work, an effort to achieve a more balanced representation across all data sources in the original dataset should be made. Detailed analysis of isolated categories could also provide additional insight into the impact of individual features within their category.

\noindent\textbf{On-chain data diversification}. Both BTC and USDC on-chain data proved very valuable in improving the performance of forecasting models. In addition to these two cryptocurrencies, on-chain data from other targeted assets should also be investigated. We propose that future work should include on-chain data from cryptocurrencies that constitute good representatives of their market category (e.g., Ethereum for Decentralized Finance) providing more holistic insights about the cryptocurrency market.

\noindent\textbf{Impact on complex models}. We have already established that data source diversity is very important for ML models designed for the cryptocurrency market. The next step is to investigate the impact of diversity on more complex models and deep learning architectures, determining whether this diversity is beneficial or introduces unnecessary noise.

\noindent\textbf{Application in finance}. The outcome of this research should be used as the baseline for building predictive models for the cryptocurrency markets. Feature engineering techniques could also help discover valuable relationships between data categories, which might further enhance the predictive capabilities of such models. In future work, we plan to use the findings and datasets of this research to build novel portfolio optimization techniques that are resilient to the highly dynamic and uncertain nature of this market. 

\section{Conclusion}

In this study, we have explored the significant role that data source diversity plays in enhancing the predictive performance of forecasting models in the cryptocurrency market. Our extensive experiments demonstrate how integrating a variety of data categories—including technical indicators, on-chain metrics, sentiment and interest metrics, traditional market indices, and macroeconomic indicators—can lead to substantial improvements in forecasting accuracy across different time horizons. We also introduced the Crypto100 index, a new market index that effectively represents the overall cryptocurrency market, and developed a novel algorithm to identify the most impactful and resilient features from a diverse set of data sources. Our findings highlight the changing importance of different data categories and individual indicators depending on the time period and prediction window, providing valuable insights for both short-term and long-term market predictions.

The results underscore the necessity of incorporating a diverse range of data sources to improve the robustness and accuracy of cryptocurrency forecasting models. Future work should focus on balancing the representation of different data categories, diversifying on-chain data, investigating the impact of data diversity on more complex models, and applying these findings to practical financial processes such as portfolio optimization. Overall, our study lays the groundwork for more accurate and resilient forecasting models, enhancing our understanding of the dynamic and volatile cryptocurrency market.%\vspace{-.1em}

%\clearpage

\bibliographystyle{ACM-Reference-Format}
\bibliography{sample}


\begin{thebibliography}{25}

%%% the Web address.

\ifx \showCODEN    \undefined \def \showCODEN     #1{\unskip}     \fi
\ifx \showDOI      \undefined \def \showDOI       #1{#1}\fi
\ifx \showISBNx    \undefined \def \showISBNx     #1{\unskip}     \fi
\ifx \showISBNxiii \undefined \def \showISBNxiii  #1{\unskip}     \fi
\ifx \showISSN     \undefined \def \showISSN      #1{\unskip}     \fi
\ifx \showLCCN     \undefined \def \showLCCN      #1{\unskip}     \fi
\ifx \shownote     \undefined \def \shownote      #1{#1}          \fi
\ifx \showarticletitle \undefined \def \showarticletitle #1{#1}   \fi
\ifx \showURL      \undefined \def \showURL       {\relax}        \fi
% invisible to TeX
\providecommand\bibfield[2]{#2}
\providecommand\bibinfo[2]{#2}
\providecommand\natexlab[1]{#1}
\providecommand\showeprint[2][]{arXiv:#2}

\bibitem[\protect\citeauthoryear{??}{coi}{2024}]%
        {coingecko_2024}
 \bibinfo{year}{2024}\natexlab{}.
\newblock \bibinfo{title}{Cryptocurrency Categories}.
\newblock
  \bibinfo{howpublished}{\url{https://www.coingecko.com/en/categories}}.
\newblock
\newblock
\shownote{Accessed: 2024-05-14.}


\bibitem[\protect\citeauthoryear{Alonso-Monsalve, Su{\'a}rez-Cetrulo,
  Cervantes, and Quintana}{Alonso-Monsalve et~al\mbox{.}}{2020}]%
        {alonso2020convolution}
\bibfield{author}{\bibinfo{person}{Sa{\'u}l Alonso-Monsalve},
  \bibinfo{person}{Andr{\'e}s~L Su{\'a}rez-Cetrulo}, \bibinfo{person}{Alejandro
  Cervantes}, {and} \bibinfo{person}{David Quintana}.}
  \bibinfo{year}{2020}\natexlab{}.
\newblock \showarticletitle{Convolution on neural networks for high-frequency
  trend prediction of cryptocurrency exchange rates using technical
  indicators}.
\newblock \bibinfo{journal}{\emph{Expert Systems with Applications}}
  \bibinfo{volume}{149} (\bibinfo{year}{2020}), \bibinfo{pages}{113250}.
\newblock


\bibitem[\protect\citeauthoryear{Bank}{Bank}{2024}]%
        {ecb_dataset}
\bibfield{author}{\bibinfo{person}{European~Central Bank}.}
  \bibinfo{year}{2024}\natexlab{}.
\newblock \bibinfo{title}{Harmonised Index of Consumer Prices (HICP)}.
\newblock
  \bibinfo{howpublished}{\url{https://data.ecb.europa.eu/data/datasets/ICP/ICP.M.U2.N.000000.4.ANR}}.
\newblock


\bibitem[\protect\citeauthoryear{Casella and Paletto}{Casella and
  Paletto}{2023}]%
        {casella2023predicting}
\bibfield{author}{\bibinfo{person}{Bruno Casella} {and}
  \bibinfo{person}{Lorenzo Paletto}.} \bibinfo{year}{2023}\natexlab{}.
\newblock \showarticletitle{Predicting Cryptocurrencies Market Phases through
  On-Chain Data Long-Term Forecasting}. In \bibinfo{booktitle}{\emph{2023 IEEE
  International Conference on Blockchain and Cryptocurrency (ICBC)}}. IEEE,
  \bibinfo{pages}{1--4}.
\newblock


\bibitem[\protect\citeauthoryear{Catalini, de~Gortari, and Shah}{Catalini
  et~al\mbox{.}}{2022}]%
        {catalini2022some}
\bibfield{author}{\bibinfo{person}{Christian Catalini}, \bibinfo{person}{Alonso
  de Gortari}, {and} \bibinfo{person}{Nihar Shah}.}
  \bibinfo{year}{2022}\natexlab{}.
\newblock \showarticletitle{Some simple economics of stablecoins}.
\newblock \bibinfo{journal}{\emph{Annual Review of Financial Economics}}
  \bibinfo{volume}{14} (\bibinfo{year}{2022}), \bibinfo{pages}{117--135}.
\newblock


\bibitem[\protect\citeauthoryear{Chicco, Oneto, and Tavazzi}{Chicco
  et~al\mbox{.}}{2022}]%
        {chicco2022eleven}
\bibfield{author}{\bibinfo{person}{Davide Chicco}, \bibinfo{person}{Luca
  Oneto}, {and} \bibinfo{person}{Erica Tavazzi}.}
  \bibinfo{year}{2022}\natexlab{}.
\newblock \showarticletitle{Eleven quick tips for data cleaning and feature
  engineering}.
\newblock \bibinfo{journal}{\emph{PLOS Computational Biology}}
  \bibinfo{volume}{18}, \bibinfo{number}{12} (\bibinfo{year}{2022}).
\newblock


\bibitem[\protect\citeauthoryear{CoinMetrics}{CoinMetrics}{2024}]%
        {coinmetrics}
\bibfield{author}{\bibinfo{person}{CoinMetrics}.}
  \bibinfo{year}{2024}\natexlab{}.
\newblock \bibinfo{title}{CoinMetrics: Crypto Intelligence for the Future of
  Finance}.
\newblock \bibinfo{howpublished}{\url{https://coinmetrics.io/}}.
\newblock
\newblock
\shownote{Accessed: 2024-06-11.}


\bibitem[\protect\citeauthoryear{{Crypto.com}}{{Crypto.com}}{2024}]%
        {CryptoCom2023}
\bibfield{author}{\bibinfo{person}{{Crypto.com}}.}
  \bibinfo{year}{2024}\natexlab{}.
\newblock \bibinfo{booktitle}{\emph{Crypto Market Sizing 2023}}.
\newblock \bibinfo{type}{{T}echnical {R}eport}.
  \bibinfo{institution}{Crypto.com}.
\newblock
\urldef\tempurl%
\url{https://contenthub-static.crypto.com/wp_media/2024/01/Crypto-Market-Sizing-2023.pdf}
\showURL{%
\tempurl}
\newblock
\shownote{Accessed: 2024-05-03.}


\bibitem[\protect\citeauthoryear{Fang, Ventre, Basios, Kanthan, Martinez-Rego,
  Wu, and Li}{Fang et~al\mbox{.}}{2022}]%
        {fang2022cryptocurrency}
\bibfield{author}{\bibinfo{person}{Fan Fang}, \bibinfo{person}{Carmine Ventre},
  \bibinfo{person}{Michail Basios}, \bibinfo{person}{Leslie Kanthan},
  \bibinfo{person}{David Martinez-Rego}, \bibinfo{person}{Fan Wu}, {and}
  \bibinfo{person}{Lingbo Li}.} \bibinfo{year}{2022}\natexlab{}.
\newblock \showarticletitle{Cryptocurrency trading: a comprehensive survey}.
\newblock \bibinfo{journal}{\emph{Financial Innovation}} \bibinfo{volume}{8},
  \bibinfo{number}{1} (\bibinfo{year}{2022}), \bibinfo{pages}{13}.
\newblock


\bibitem[\protect\citeauthoryear{Griffin and Shams}{Griffin and Shams}{2020}]%
        {griffin2020bitcoin}
\bibfield{author}{\bibinfo{person}{John~M Griffin} {and} \bibinfo{person}{Amin
  Shams}.} \bibinfo{year}{2020}\natexlab{}.
\newblock \showarticletitle{Is Bitcoin really untethered?}
\newblock \bibinfo{journal}{\emph{The Journal of Finance}}
  \bibinfo{volume}{75}, \bibinfo{number}{4} (\bibinfo{year}{2020}),
  \bibinfo{pages}{1913--1964}.
\newblock


\bibitem[\protect\citeauthoryear{Invesco}{Invesco}{2024}]%
        {invesco_uup_etf}
\bibfield{author}{\bibinfo{person}{Invesco}.} \bibinfo{year}{2024}\natexlab{}.
\newblock \bibinfo{title}{Invesco DB US Dollar Index Bullish Fund (UUP)
  Profile}.
\newblock
  \bibinfo{howpublished}{\url{https://www.invesco.com/us/financial-products/etfs/product-detail?audienceType=Investor&ticker=UUP}}.
\newblock
\newblock
\shownote{Accessed: 2024-06-11.}


\bibitem[\protect\citeauthoryear{Investopedia}{Investopedia}{2024}]%
        {investopediasp500calculation}
\bibfield{author}{\bibinfo{person}{Investopedia}.}
  \bibinfo{year}{2024}\natexlab{}.
\newblock \bibinfo{title}{How Is the S\&P 500 Index Calculated?}
\newblock
  \bibinfo{howpublished}{\url{https://www.investopedia.com/ask/answers/05/sp500calculation.asp}}.
\newblock
\newblock
\shownote{Accessed: 2024-06-11.}


\bibitem[\protect\citeauthoryear{Kim, Bock, and Lee}{Kim et~al\mbox{.}}{2021}]%
        {kim2021predicting}
\bibfield{author}{\bibinfo{person}{Han-Min Kim}, \bibinfo{person}{Gee-Woo
  Bock}, {and} \bibinfo{person}{Gunwoong Lee}.}
  \bibinfo{year}{2021}\natexlab{}.
\newblock \showarticletitle{Predicting Ethereum prices with machine learning
  based on Blockchain information}.
\newblock \bibinfo{journal}{\emph{Expert Systems with Applications}}
  \bibinfo{volume}{184} (\bibinfo{year}{2021}), \bibinfo{pages}{115480}.
\newblock


\bibitem[\protect\citeauthoryear{LunarCrush}{LunarCrush}{2024}]%
        {lunarcrush}
\bibfield{author}{\bibinfo{person}{LunarCrush}.}
  \bibinfo{year}{2024}\natexlab{}.
\newblock \bibinfo{title}{LunarCrush: Social Intelligence for Crypto}.
\newblock \bibinfo{howpublished}{\url{https://lunarcrush.com/}}.
\newblock
\newblock
\shownote{Accessed: 2024-06-11.}


\bibitem[\protect\citeauthoryear{Lundberg and Lee}{Lundberg and Lee}{2017}]%
        {lundberg2017unified}
\bibfield{author}{\bibinfo{person}{Scott~M Lundberg} {and}
  \bibinfo{person}{Su-In Lee}.} \bibinfo{year}{2017}\natexlab{}.
\newblock \showarticletitle{A unified approach to interpreting model
  predictions}.
\newblock \bibinfo{journal}{\emph{Advances in neural information processing
  systems}}  \bibinfo{volume}{30} (\bibinfo{year}{2017}).
\newblock


\bibitem[\protect\citeauthoryear{Ma, Ahmad, Liu, and Wang}{Ma
  et~al\mbox{.}}{2020}]%
        {ma2020portfolio}
\bibfield{author}{\bibinfo{person}{Yechi Ma}, \bibinfo{person}{Ferhana Ahmad},
  \bibinfo{person}{Miao Liu}, {and} \bibinfo{person}{Zilong Wang}.}
  \bibinfo{year}{2020}\natexlab{}.
\newblock \showarticletitle{Portfolio optimization in the era of digital
  financialization using cryptocurrencies}.
\newblock \bibinfo{journal}{\emph{Technological forecasting and social change}}
   \bibinfo{volume}{161} (\bibinfo{year}{2020}), \bibinfo{pages}{120265}.
\newblock


\bibitem[\protect\citeauthoryear{Mokni, El~Montasser, Ajmi, and Bouri}{Mokni
  et~al\mbox{.}}{2024}]%
        {mokni2024efficiency}
\bibfield{author}{\bibinfo{person}{Khaled Mokni}, \bibinfo{person}{Ghassen
  El~Montasser}, \bibinfo{person}{Ahdi~Noomen Ajmi}, {and}
  \bibinfo{person}{Elie Bouri}.} \bibinfo{year}{2024}\natexlab{}.
\newblock \showarticletitle{On the efficiency and its drivers in the
  cryptocurrency market: the case of Bitcoin and Ethereum}.
\newblock \bibinfo{journal}{\emph{Financial Innovation}} \bibinfo{volume}{10},
  \bibinfo{number}{1} (\bibinfo{year}{2024}), \bibinfo{pages}{39}.
\newblock


\bibitem[\protect\citeauthoryear{Nakamoto}{Nakamoto}{[n.d.]}]%
        {satoshi}
\bibfield{author}{\bibinfo{person}{Satoshi Nakamoto}.}
  \bibinfo{year}{[n.d.]}\natexlab{}.
\newblock \showarticletitle{Bitcoin: A Peer-to-Peer Electronic Cash System}.
\newblock  (\bibinfo{year}{[n.\,d.]}).
\newblock
\urldef\tempurl%
\url{www.bitcoin.org}
\showURL{%
\tempurl}


\bibitem[\protect\citeauthoryear{Ortu, Uras, Conversano, Bartolucci, and
  Destefanis}{Ortu et~al\mbox{.}}{2022}]%
        {ortu2022technical}
\bibfield{author}{\bibinfo{person}{Marco Ortu}, \bibinfo{person}{Nicola Uras},
  \bibinfo{person}{Claudio Conversano}, \bibinfo{person}{Silvia Bartolucci},
  {and} \bibinfo{person}{Giuseppe Destefanis}.}
  \bibinfo{year}{2022}\natexlab{}.
\newblock \showarticletitle{On technical trading and social media indicators
  for cryptocurrency price classification through deep learning}.
\newblock \bibinfo{journal}{\emph{Expert Systems with Applications}}
  \bibinfo{volume}{198} (\bibinfo{year}{2022}), \bibinfo{pages}{116804}.
\newblock


\bibitem[\protect\citeauthoryear{Ren, Ma, Kong, Baltas, and Zureigat}{Ren
  et~al\mbox{.}}{2022}]%
        {ren2022past}
\bibfield{author}{\bibinfo{person}{Yi-Shuai Ren}, \bibinfo{person}{Chao-Qun
  Ma}, \bibinfo{person}{Xiao-Lin Kong}, \bibinfo{person}{Konstantinos Baltas},
  {and} \bibinfo{person}{Qasim Zureigat}.} \bibinfo{year}{2022}\natexlab{}.
\newblock \showarticletitle{Past, present, and future of the application of
  machine learning in cryptocurrency research}.
\newblock \bibinfo{journal}{\emph{Research in Intl Business and Finance}}
  \bibinfo{volume}{63} (\bibinfo{year}{2022}).
\newblock


\bibitem[\protect\citeauthoryear{Schwartz}{Schwartz}{2009}]%
        {schwartz2009origins}
\bibfield{author}{\bibinfo{person}{Anna~J Schwartz}.}
  \bibinfo{year}{2009}\natexlab{}.
\newblock \showarticletitle{Origins of the financial market crisis of 2008}.
\newblock \bibinfo{journal}{\emph{Cato J.}}  \bibinfo{volume}{29}
  (\bibinfo{year}{2009}), \bibinfo{pages}{19}.
\newblock


\bibitem[\protect\citeauthoryear{Scikit-learn}{Scikit-learn}{2024}]%
        {scikit-learn_grid_search}
\bibfield{author}{\bibinfo{person}{Scikit-learn}.}
  \bibinfo{year}{2024}\natexlab{}.
\newblock \bibinfo{title}{Grid Search}.
\newblock
  \bibinfo{howpublished}{\url{https://scikit-learn.org/stable/modules/grid_search.html}}.
\newblock
\newblock
\shownote{Accessed: 2024-06-09.}


\bibitem[\protect\citeauthoryear{Sebasti{\~a}o and Godinho}{Sebasti{\~a}o and
  Godinho}{2021}]%
        {sebastiao2021forecasting}
\bibfield{author}{\bibinfo{person}{Helder Sebasti{\~a}o} {and}
  \bibinfo{person}{Pedro Godinho}.} \bibinfo{year}{2021}\natexlab{}.
\newblock \showarticletitle{Forecasting and trading cryptocurrencies with
  machine learning under changing market conditions}.
\newblock \bibinfo{journal}{\emph{Financial Innovation}}  \bibinfo{volume}{7}
  (\bibinfo{year}{2021}), \bibinfo{pages}{1--30}.
\newblock


\bibitem[\protect\citeauthoryear{Uncertainty}{Uncertainty}{2024}]%
        {policy_uncertainty}
\bibfield{author}{\bibinfo{person}{Economic~Policy Uncertainty}.}
  \bibinfo{year}{2024}\natexlab{}.
\newblock \bibinfo{title}{Measuring Economic Policy Uncertainty}.
\newblock \bibinfo{howpublished}{\url{https://www.policyuncertainty.com}}.
\newblock
\newblock
\shownote{Accessed: 2024-06-11.}


\bibitem[\protect\citeauthoryear{Vanguard}{Vanguard}{2024}]%
        {vanguard_bsv_etf}
\bibfield{author}{\bibinfo{person}{Vanguard}.} \bibinfo{year}{2024}\natexlab{}.
\newblock \bibinfo{title}{Vanguard Short-Term Bond ETF (BSV) Profile}.
\newblock
  \bibinfo{howpublished}{\url{https://investor.vanguard.com/investment-products/etfs/profile/bsv}}.
\newblock
\newblock
\shownote{Accessed: 2024-06-11.}


\end{thebibliography}

\end{document}